\documentclass[aps]{revtex4}
\usepackage{epsfig,graphics,amssymb,amsmath,subeqnarray}
\def\ss{\boldsymbol\sigma}
\def\d{{\rm d}}
\def\p{\partial}
\def\V{{\bf V}}
\def\U{{\bf U}}
\def\w{{\bf w}}
\def\u{{\bf u}}

\def\k{{\bf k}}
\def\r{{\bf r}}

\def\g{{\bf g}}
\def\G{{\bf G} }

\def\t{\tilde}
\def\F{{\cal F}}
\def\I{{\cal I}}
\def\J{{\cal J}}
\def\br{{\bf r}}
\def\F1{{\cal F}^-1}
\def\kn{{\rm Kn}}

\newcommand{\pd}[2]{\frac{\partial #1}{\partial #2}}

\newcommand{\oo}[1]{{\cal O}\left( #1 \right)}    \def\.{^{(.)}}
\def\be{\begin{equation}}
\def\ee{\end{equation}}
\begin{document}

\title{\textbf{Brownian motion near a partial-slip boundary: \\ A local probe of the no-slip condition}}
\author{Eric Lauga}
\affiliation{Division of Engineering and Applied Sciences,
Harvard University, Cambridge MA 02138.}
\author{Todd M. Squires}
\affiliation{Department of Physics and Department of Applied and
Computational Mathematics, California Institute of Technology
Pasadena, CA 91125.}
\date{\today}

\begin{abstract}
Motivated by experimental evidence of violations of the no-slip boundary condition for liquid flow in micron-scale geometries, we propose a simple, complementary experimental technique that has certain advantages over previous studies.  Instead of relying on externally-induced flow or probe motion, we suggest that colloidal diffusivity near solid surfaces contains signatures of the degree of fluid slip exhibited on those surfaces.  To investigate, we calculate the image system for point forces (Stokeslets) oriented perpendicular and parallel to a surface with a finite slip length, analogous to Blake's solution for a Stokeslet near a no-slip wall.  Notably, the image system for the point source and perpendicular Stokeslet contain the same singularities as Blake's solution; however, each is distributed along a line with a magnitude that decays exponentially over the slip length.  The image system for the parallel Stokeslet involves a larger set of fundamental singularities, whose magnitude does not decay exponentially from the surface.  Using these image systems, we determine the wall-induced correction to the diffusivity of a small spherical particle located `far' from the wall.  We also calculate the coupled diffusivities between multiple particles near a partially-slipping wall.  Because, in general, the diffusivity depends on `local' wall conditions, patterned surfaces would allow differential measurements to be obtained within a single experimental cell, eliminating potential cell-to-cell variability encountered in previous experiments.  In addition to motivating the proposed experiments, our solutions for point forces and sources near a partial-slip wall will be useful for boundary integral calculations in slip systems.
\end{abstract}
\maketitle

\section{Introduction}

Recent reports of an apparent breakdown of the no-slip boundary condition for liquid flows in small geometries provide an exciting and surprising opportunity to revisit one of the most fundamental questions in hydrodynamics. It has become textbook knowledge that, in the framework of continuum mechanics, the velocity of a viscous fluid at a solid boundary is equal to that of the solid.  If the solid is at rest, the adjacent fluid must also be at rest.  Although it can not be derived from first principles, decades of agreement with experiments has led to a consensus that the no-slip boundary condition is indeed correct for the fluid/solid boundary \cite{goldstein38}.

The simplest and most natural violation of the no-slip condition would involve a surface slip velocity that varies in proportion to the local shear rate, written in the case of a flat surface as
\begin{equation}\label{navier}
{\bf u}_\parallel = \lambda \frac{\partial {\bf u}_\parallel
}{\partial n},\quad u_\perp = 0.
\end{equation}
This condition naturally introduces a new length scale, $\lambda$, called the slip length.  In gases, the slip length is related to the mean free path, $\lambda_f$, where non-continuum effects become important \cite{maxwell79}.  The analogous picture does not appear to hold for liquids, however.  Liquid molecules are in constant collision, and any analogous mean free path would be of molecular order, significantly smaller than recent experiments suggest (discussed below).  Regardless of the discrepancy between physical origins of apparent slip in liquids and gases, the effects of slip are expected to become important when the experimental length scale $h$ is of the same order as the slip length $\lambda$.  Therefore, by analogy with gaseous slip flows, we will use an effective Knudsen number,
\be
\kn = \frac{\lambda}{h}
\ee
to describe flows near a partial-slip surface.  Obviously, we expect slip effects to play a significant role when $\kn \gtrsim \oo{1}$.

In the past half-century, the no-slip condition saw only occasional challenges \cite{churaev84,schnell56}.  More recently, however, various experimental systems have probed liquid flows on small enough length scales $h$ that Kn may no longer be small, allowing a more thorough and sustained re-investigation of the no-slip boundary condition. These experimental techniques differ in the way that flow is created and slip is measured, and fall into five primary categories, reviewed in \cite{lauga05}.  (i) One can measure the relationship between flow rate and pressure drop in capillaries or microchannels \cite{cheng02,choi03}, which depends on Kn.  (ii) One can measure the force required for squeeze flows in long and narrow geometries such as are found in the surface force apparatus (SFA) or atomic force microscope (AFM) \cite{baudry01,bonaccurso02,bonaccurso03,cottinbizonne02,craig01b,zhu01,sun02,zhu02d,neto03,cho04,henry04,cottin-bizonne05}.  Slip reduces the viscous resistance, and quasi-steady probe motion is assumed.  (iii) One can measure pressure-driven velocity profiles in a capillary or microchannel using small particles as passive tracers \cite{tretheway02,joseph05}.  Tracers that are sufficiently small, uncharged, and far from the wall should faithfully reproduce the fluid velocity, although the high diffusivity of small probes requires averaging techniques.  (iv) One can measure an externally-driven flow near a wall using fluorescence correlation spectroscopy with labelled molecular tracers \cite{lumma03}.  (v)  One can use near field laser velocimetry, wherein evanescent optical waves (exponentially localized to a small region near a wall) are used to measure the velocity of photobleached molecular probes in an externally-driven flow \cite{pit00,hervet03}.

There are now many published reports of apparent violations of the no-slip condition, both experimental \cite{boehnke99,kiseleva99,baudry01,bonaccurso02,bonaccurso03,cheikh03,cheng02,choi03,cottinbizonne02,craig01b,tretheway02,zhu01,zhu02d,lumma03,pit00,joseph05,sun02,neto03,cho04,henry04,cottin-bizonne05,hervet03} and theoretical \cite{thompson90,thompson97,barrat99b}.  Apparent slip has been measured over surfaces that are completely wetting \cite{pit00,bonaccurso02,bonaccurso03}, partially wetting \cite{zhu01,craig01b}, and non-wetting \cite{tretheway02,barrat99b,zhu01,baudry01,choi03,cottinbizonne02}.  Roughness has been predicted and measured to decrease slip \cite{richardson73,jansons88,zhu02d,pit00}, although in some cases roughness appears to increase slip \cite{cottinbizonne03,bonaccurso03}.  In some measurements and simulations, the slip length appears to be  independent of shear rate \cite{barrat99b,baudry01,pit00,cottinbizonne02}, whereas in others it depends upon shear rate  \cite{thompson97,choi03,zhu01,bonaccurso02,bonaccurso03,craig01b}.  Moreover, apparent slip lengths ranging from nanometers \cite{bonaccurso02} to microns \cite{zhu01} have been been reported.  It is thus reasonable to conclude that no consensus has been reached concerning the existence and physical origin of fluid/solid slip, and the physical factors that influence it.

The large variability in the published results could be due in part to the variety of experimental techniques employed. After all, physical mechanisms other than liquid/solid slip can resemble apparent slip in experiments \cite{lauga04a,lauga03,lauga04b,degennes02,tretheway04}, and could lead to incorrect conclusions as to the nature of the actual solid/liquid interface.  Different experimental techniques are susceptible to these effects to different degrees.  Additionally, all require an externally-forced flow or motion, which introduces an additional source of experimental uncertainty.  Furthermore, multiple experimental cells are typically required to probe different solid/liquid surfaces.  Finally, many experiments involve averaging -- over the length of a capillary, the area of an SFA, or the diffusive motion of tracers.

In this paper, we propose a complementary technique to probe the nature of the liquid/solid boundary that is largely immune to the issues raised above.  
The idea is to measure the influence of the wall on the Brownian motion of suspended tracers. A spherical particle of radius $a$, far from the wall, diffuses with a bulk diffusivity  $D_0=k_BT/6\pi\mu a$.  When a wall is located a distance $h$ from the particle, particle diffusivity is affected in a manner that depends on the nature of the surface.  A no-slip wall (${\rm Kn}=0$) gives corrections to the perpendicular ($D^\perp$) and parallel ($D^\parallel$) diffusion coefficients,
\begin{equation}\label{noslipdiff}
D^\perp=D_0\left(1-\frac{9a}{8h}\right),\quad
D^\parallel=D_0\left(1-\frac{9a}{16h}\right),
\end{equation}
with errors of order ${\cal O}(a^3/h^3)$ \cite{happel83}. If, however, the surface is perfectly slipping ({\it i.e.}  sustains no shear stress, or ${\rm Kn}=\infty$), the particle diffusivities are given by
\begin{equation}\label{slipdiff}
D^\perp=D_0\left(1-\frac{3a}{4h}\right),\quad
D^\parallel=D_0\left(1+\frac{3a}{8h}\right)\cdot
\end{equation}
Finite values of the slip length (or Kn) should interpolate between these two limits.  We note in particular that the parallel diffusivity goes from being wall hindered for ${\rm Kn}\ll 1$ to enhanced for ${\rm Kn}\gg 1$.  Naturally, a knowledge of the relation between slip length and diffusivity would allow the slip length of a solid/fluid interface to be inferred from measured diffusivity of nearby particles.  No external flow is required, and walls with patterned wettability allow various surfaces to be probed within a single experimental cell, which would allow differential measurements that are free of the uncertainties due to cell-to-cell variability.  This builds upon an idea that was first pursued by Alm\'eras {\sl et al.} \cite{almeras00}, who characterized the influence of wettability and slip on the parallel diffusion coefficient of a small particle between two walls, as involved in molecular diffusion under confinement.  

In this work, we calculate fundamental solutions for Stokes flows near a single partial-slip wall.  Our results give an explicit relationship between solid/liquid slip and colloidal diffusivity, as well as expressions for the flow fields themselves.  In addition to aiding in intuition for partial-slip systems, the flow fields we calculate will be useful for boundary integral calculations in partial-slip systems.  Additionally, we explore the feasibility of measuring the effect of a partial-slip wall upon colloidal diffusivity as a means of measuring the wall slip itself.  Recent years have seen precise experimental measurements of colloidal diffusivity near walls and/or other colloids.  Corrections of order $a/h$ can be accurately measured, and excellent agreement has been found with theory \cite{dufresne00,dufresne01,crocker97}.  

This technique has various advantages.  First, it does not require an external flow and therefore alleviates the experimental difficulties associated with precise flow manipulation. As a consequence, the experiment can be performed in a closed cell, and thus avoid contamination by impurities.  Consequently, the liquid can be degassed or put under variable pressure to probe the influence of adsorbed nanobubbles, as discussed below.  Second, our method does not average over different experiments, sample volume or apparatus size but instead makes use of a single colloidal probe.  Third, multiple solid/liquid interfaces can be probed within a single experimental cell by using deliberately patterned surfaces.  This would allow differential measurements to be performed, and possibly to track surface-attached nanobubbles \cite{ishida00,tyrrell01,tyrrell02}.  

The paper is organized as follows. In \S\ref{Image}, we consider the effects of a partial-slip wall on the two main fundamental singularities of Stokes flow -- the point force (Stokeslet) and the point source.  In a manner analogous to Blake's image system for a no-slip wall \cite{blake71}, we interpret the wall's contribution in terms of a series of image singularities.  This is the central result of our work.  In \S \ref{Influence}, we use this solution to provide an analytical formula for the influence of a partial-slip wall on the diffusivity of a small spherical particle.   In \S\ref{correlated}, we consider the coupled mobilities/diffusivities of two small particles, and propose alternate experimental tests for slip to complement those in \S \ref{Influence}. As the calculations themselves are somewhat laborious, we relegate the details to appendices, and save the main body of the text for key results and discussion.

\section{Image systems near a partial slip surface}
\label{Image}

Analogous to point charges and point masses in electrostatics and gravitation, flow fields associated with fundamental singularities are useful in treating Stokes flows \cite{pozrikidis92}.  There are two families of fundamental singularities in Stokes flows -- (i) the point source/sink and their derivatives, which correspond to irrotational potential flow and are entirely analogous to electrostatic fields, and (ii) the point force (Stokeslet) and its derivatives, whose flow fields are viscous and rotational.  These fundamental singularities aid in intuition for viscous flows, and in providing approximate and asymptotic solutions.  Furthermore, they form the basis for boundary integral techniques in Stokes flow calculations in more complicated geometries, where flow and pressure fields are computed by solving for surface distributions of fundamental singularities \cite{pozrikidis92}.

In a classic paper, Blake \cite{blake71} interpreted the flow field due to a Stokeslet near a no-slip surface in terms of a system of image singularities, located on the opposite side of the wall.  Blake's image system consists of an  equal but opposite Stokeslet, a Stokeslet dipole ({\it i.e.}, force-dipole) and a source dipole (potential dipole).  A perfectly slipping surface ($\kn = \infty$), has a simpler image system:  A single Stokeslet of equal magnitude and symmetric direction, as in Fig.~\ref{firstimagefig}.  The image system for a partial-slip surface ($0<\kn<\infty$) is more complicated, and is the subject of the following analysis.  In what follows, we present the complete image system for a Stokeslet and a point source near a planar partial-slip boundary.  Higher order (multipolar) singularities can be derived from these two by differentiation, although subtleties exist (discussed below).  Because our calculation is analogous to Blake's, but algebraically more involved, we save the details for Appendices \ref{solution},  \ref{solve_perpendicular}, \ref{solve_parallel} and \ref{Fourier}.  

\subsection{Set up and boundary conditions}
\label{setup}

\begin{figure}[t]
\centering
\includegraphics[width=.7\textwidth]{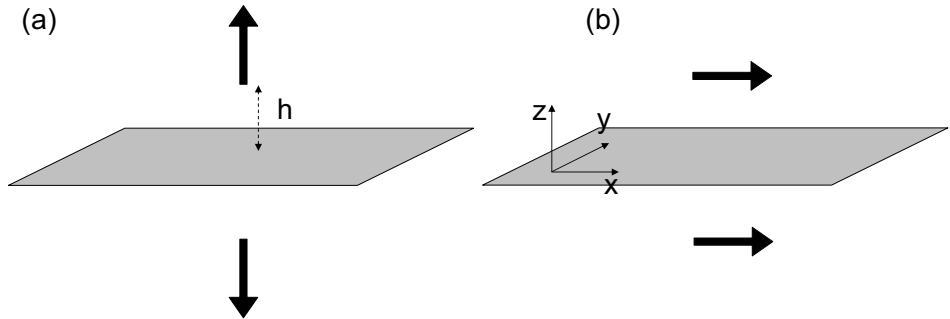}
\caption{First image of the Stokeslet; (a): Stokeslet ($F,h$)
perpendicular to the surface; the first image is the Stokeslet
($-F,-h$); (b): Stokeslet ($F,h$) parallel to the surface; the
first image is the Stokeslet ($F,-h$). Note that these would be
the complete image systems if the surface was perfectly slipping
($\kn = \infty$).} \label{firstimagefig}
\end{figure}

We choose the $x-y$ plane to lie along the solid wall, with the $z$ coordinate directed perpendicular to the surface, and consider a Stokeslet of strength ${\bf F}$ located at $(x,y,z)=(0,0,h)$. The velocity field, ${\bf u}$, satisfies the incompressible Stokes equations
\begin{equation}\label{stokes}
\mu\nabla^2 {\bf u}=\nabla p,\quad \nabla\cdot{\bf u}={\bf 0},
\end{equation}
subject to partial slip boundary conditions  (Eq.~\ref{navier}).
We decompose {\bf u} into three components,
\begin{equation}\label{decomp}
{\bf u}={\bf U}+{\bf V}+{\bf w},
\end{equation}
where ${\bf U}$ is the flow field due to the Stokeslet itself, ${\bf V}$ is the flow field due to the primary image  Stokeslet of strength $\tilde {\bf F}$ located at $(x,y,z)=(0,0,-h)$ (Fig. \ref{firstimagefig}), and ${\bf w}$ is an as yet unknown velocity field that solves Eq.~\eqref{stokes}. The Green's function for the Stokeslet is given by \cite{blake71,pozrikidis92}
\begin{equation}
{\bf G}^S({\bf r})=\frac{1}{8\pi\mu}\left(\frac{\bf 1}{ r} + \frac{\bf r r}{{ r}^3}\right)\cdot
\end{equation}

Since the wall breaks the isotropy of the particle mobility, we consider the perpendicular ($^\perp$) and parallel ($^\parallel$) Stokeslets separately, giving velocity fields
\begin{subeqnarray}
\U^\perp & = & F\, {\bf G}^S_{z}({\bf r})
\slabel{Uperp} \\
\U^\parallel & = & F\,{\bf G}^S_{x}({\bf r})
,\slabel{Uparallel}
\end{subeqnarray}
with ${\bf r}=(x,y,z-h)$. Here we have introduced the notation
\be
{\bf G}^S_{z}({\bf r}) = {\bf G}^S ({\bf r-r}_0)\cdot {\bf e}_z,
\ee
where ${\bf r}_0$ is the location of the Stokeslet, ${\bf r}$ is the observation point, and 
the subscript indicates the direction of the force.  In what follows, we will also use the following notation for higher-order singularities
\begin{subeqnarray}
\left.{\bf G}^{SD}_{z;x}({\bf r}) = \pd{}{x_0}\left\{{\bf G}^S({\bf r-r}_0) \cdot {\bf e}_z\right\}\right|_{{\bf r}_0={\bf 0}}\equiv-\pd{}{x}\left\{{\bf G}^S({\bf r}) \cdot {\bf e}_z\right\}\\
\left.{\bf G}^{SQ}_{z;xy}({\bf r}) = \frac{\partial^2}{\partial x_0 \partial y_0}\left\{{\bf G}^S({\bf r-r}_0) \cdot {\bf e}_z\right\}\right|_{{\bf r}_0={\bf 0}}\equiv \frac{\partial^2}{\partial x \partial y}\left\{{\bf G}^S({\bf r}) \cdot {\bf e}_z\right\}.
\end{subeqnarray}
Here ${\bf G}^{SD}$ represents a Stokeslet doublet, ${\bf G}^{SQ}$ a Stokeslet quadrupole, and so on.  Note that derivatives are taken with respect to the singularity location (rather than the observation point), one derivative for each coordinate following the semicolon.

For the primary image Stokeslets located at $z=-h$, we pick $\tilde {\bf F}$ as in Fig.~\ref{firstimagefig} to enforce the no-flux condition at the wall, giving velocity fields 
\begin{subeqnarray}
{\bf V}^\perp & = & -F\, {\bf G}^S_z(\bar{\bf r})
\quad 
\slabel{Vperp} \\
\V^\parallel & = & F\,{\bf G}_x^S(\bar {\bf r}) 
\slabel{Vparallel}
\end{subeqnarray}
where $\bar {\bf  r}=(x,y,z+h)$.

Enforcing the partial-slip boundary condition (Eq.~\ref{navier}) at the surface $z=0$ imposes boundary conditions on ${\bf w}$,
\begin{equation}\label{newbound}
\left(1-\lambda \frac{ \partial }{\partial z}\right)\w_\parallel = -2\U_\parallel,\quad w_3=0.
\end{equation}
We use Fourier transforms in the $x$- and $y$-directions and the general solution to Stokes equations, as given in Appendix \ref{solution}. The amplitudes of the Fourier components are determined by enforcing the slip boundary condition \eqref{newbound}; for details see Appendix \ref{solve_perpendicular} for the perpendicular case, and Appendix \ref{solve_parallel} for the parallel case.  These solutions can be used to calculate the change in the particle mobility, and therefore diffusivity.

An interesting result is that the Fourier coefficients for both cases can be related directly to the coefficients in the no-slip series, which allows the partial-slip solution to be expressed in terms of weighted integrals of the no-slip image systems.  This results in a clear physical interpretation of the partial-slip image system in terms of weighted integrals of fundamental singularities.

\subsection{Stokeslet perpendicular to slip surface}

\begin{figure}[t]
\centerline{
\includegraphics[width=.95\textwidth]{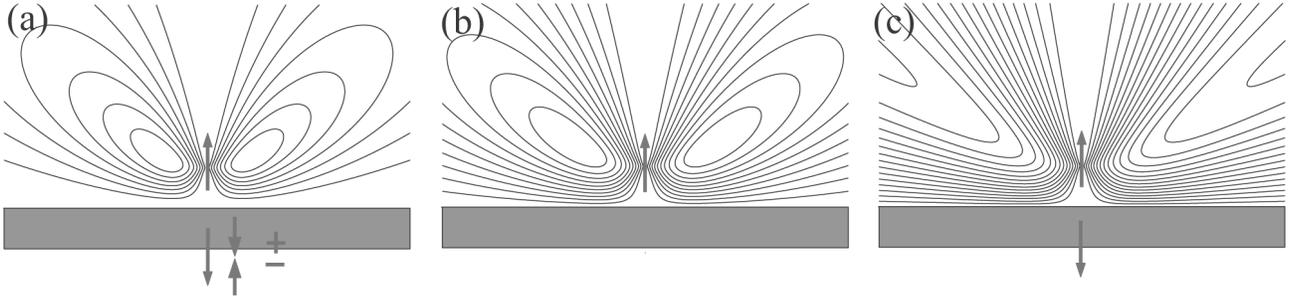}
}
\caption{Streamlines for a Stokeslet oriented perpendicular to a partial-slip wall, with (a) $\kn=0$ (Blake's solution, no-slip),  (b) $\kn = 1$, (c) $\kn=\infty$ (perfect slip). The streamlines are displayed in the plane which includes the Stokeslet and is perpendicular to the nearby surface.} \label{fig:perpfigs}
\end{figure}

As shown in Appendix \ref{solve_perpendicular}, the total velocity field for a Stokeslet oriented perpendicular to a partial-slip wall can be expressed as $\u^\perp=\U^\perp+\V^\perp+\w^\perp$, where $\U^\perp$ and $\V^\perp$ are given by Eqs.~\eqref{Uperp}
and \eqref{Vperp}, and $\w^\perp$ is given by
 \begin{equation}\label{interpretation_perp}
\w^\perp(\r,\lambda)=\frac{Fh}{ \lambda} \int_0^\infty
e^{{-s}/{2\lambda}} [(h+s)\,\G^D_z-\G^{SD}_{z;z}]({\bf  r} + (h+s){\bf e}_z)\d s,
\end{equation}
where $\G^D$ is a potential (source) dipole, defined by
\be
\left.\G^D_z = \frac{1}{8\pi\mu} \pd{}{z_0} \left(\frac{\br-\br_0}{|\br-\br_0|^3}\right)\right|_{\br_0 = {\bf 0} }\equiv -\frac{1}{8\pi\mu}\pd{}{z} \left(\frac{\br}{r^3}\right).
\ee 
Analogous formulae and notation for source quadrupoles $\G^Q$ and so on follow in a straightforward fashion.  This image system therefore represents a weighted line integral of source dipoles and Stokeslet dipoles, whose magnitude decays exponentially with distance (scaled by the slip length).  In fact, Eq. \eqref{interpretation_perp} represents a line integral of Blake's image system for no-slip walls \cite{blake74}.  The streamlines for the complete image system are displayed  in Fig.~\ref{fig:perpfigs}.

Scaling the integration variable $s$ with  $\lambda u$ gives a form for the image system,
\be
\w^\perp(\r,\lambda)= Fh \int_0^\infty
e^{{-u}/{2}} [(h+\lambda u)\,\G^D_z-\G_{z;z}^{SD}]({\bf  r} + (h+\lambda u){\bf e}_z)\d u,
\label{wperpu}
\ee
which is amenable to asymptotic analysis.  The basis for the analysis that follows is that the integrand is only appreciable when $u\lesssim\oo{1}$, and is exponentially small otherwise.

Since $z$ and $h$ play the same role in the argument of the two singularities in Eq.~\eqref{wperpu}, we can consider the limit where  $\lambda \ll (z+h)$ in the singularities in Eq.~\eqref{wperpu}. A Taylor expansion of the term in brackets gives
\begin{equation}\label{asymtot_perp}
\w^\perp(\r)\approx 2Fh[ (h+2\lambda) \G^D_z-\G_{z;z}^{SD}](\bar {\bf  r}) + 
4Fh \lambda [-(h+4\lambda)\G^Q_{zz}+\G_{z;zz}^{SQ}](\bar {\bf  r}).
\end{equation}
In the limit when the particle is farther from the wall than the slip length ($\lambda \ll h$, or $\kn \ll 1$), Eq.~\eqref{asymtot_perp} becomes
\be
\w^\perp(\r,\lambda\ll h)=2Fh 
[h\,\G_z^D-\G_{z;z}^{SD}](\bar {\bf  r}) + 4Fh\lambda
[\G_z^D-h \G_{zz}^Q+\G_{z;zz}^{SQ}](\bar {\bf  r}).
\label{wperpusmallK}
\ee
The first term is Blake's solution for a no-slip wall, and the second term represents an $\oo{\kn}$ correction to the image system due to slip. Furthermore, in the limit where $h \ll \lambda \ll z$, Eq.~\eqref{asymtot_perp} results in 
\be
\w^\perp(\r)\approx 4Fh\lambda \G_z^D(\bar {\bf  r}),
\label{wperpulargez}
\ee
which differs significantly from Blake's solution.

\subsection{Stokeslet parallel to slip surface}
\label{image_parallel} 

As shown in Appendix \ref{solve_parallel}, the total velocity field for a Stokeslet oriented parallel
to a partial-slip wall can be expressed as $\u^\parallel=\U^\parallel+\V^\parallel+\w^\parallel$, where
$\U^\parallel$ and $\V^\parallel$ are given by Eqs.~\eqref{Uparallel}
and \eqref{Vparallel}, and $\w^\parallel$ is given by
\begin{eqnarray}\label{imagepara}
\w^\parallel(\r,\lambda) & = & \frac{F}{\lambda} \int_0^\infty 
[-\G^S_x+h\G_{z;x}^{SD}-h^2\G_x^{D}]({\bf  r} + (h+s){\bf e}_z)
\,e^{{-s}/{2\lambda}}\,\d s\nonumber \\
& - & {4F\lambda} \int_0^\infty \G_{z;y}^{RD}({\bf  r} + (h+s){\bf e}_z)\,\left[e^{{-s}/{2\lambda}} - 1\right]^2\,\d s \\
& + & 4 F \lambda  \int_0^\infty [\G_x^D -h \G_{xz}^{Q} ]({\bf  r} + (h+s){\bf e}_z) \left[ 1-\left(1+\frac{s}{2\lambda}\right)\,e^{{-s}/{2\lambda}}
\right]
\,\d s, \nonumber
\end{eqnarray}
where 
$\G^{RD}_{z;y}$ is a rotlet dipole, given by (see details in Appendix \ref{solve_parallel})
\begin{equation}
\G^{RD}_{z;y}=\frac{1}{2}(\G^{SD}_{y;xy}-\G^{SD}_{x;yy}).
\end{equation}

\begin{figure}[t]
\centerline{
\includegraphics[width=.95\textwidth]{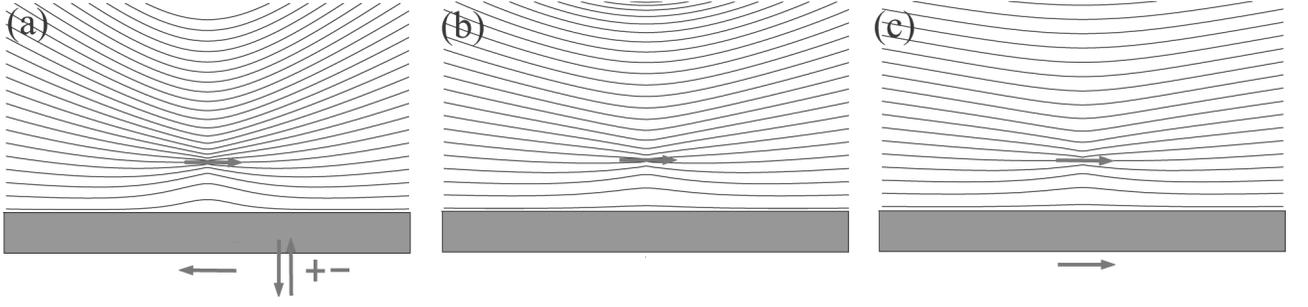}
}
\caption{Streamlines for a Stokeslet oriented parallel to a partial-slip wall, with (a) $\kn=0$ (Blake's solution, no-slip),  (b) $\kn = 1$, (c) $\kn=\infty$ (perfect slip). The streamlines are displayed in the plane which includes the Stokeslet and is perpendicular to the nearby surface.} \label{fig:parafigs}
\end{figure}

The solution, Eq.~\eqref{imagepara}, can again be interpreted as a line integral of the fundamental singularities above, but their weight does not systematically decay exponentially away from the image location. These are shown in Fig.~\ref{fig:parafigs}.  Note that, as $\lambda\to \infty$, each term in Eq.~\eqref{imagepara} goes to zero due to the rapid spatial decay of the singularities ($\G^D\sim 1/r^3$, $\G^{RD} \sim 1/r^3$, $\G^Q\sim 1/r^4$) and due to the vanishing value of the weights in Eq.~\eqref{imagepara} as $s\to 0$.

Re-scaling $s$ by $\lambda$ in Eq. gives
\begin{eqnarray}\label{imagepara_scaled}
\w^\parallel(\r,\lambda) & = & {F} \int_0^\infty 
[-\G^S_x+h\G^{SD}_{z;x}-h^2\G_x^{D}]({\bf  r} + (h+\lambda u){\bf e}_z)
\,e^{{-u}/{2}}\,\d u \nonumber \\
& - & {4F\lambda^2} \int_0^\infty \G^{RD}_{z;y}({\bf  r} + (h+\lambda u){\bf e}_z)\,\left[e^{{-u}/{2}} - 1\right]^2\,\d u \\
& + & 4 F \lambda^2  \int_0^\infty [\G^D_x -h \G^{Q}_{xz} ]({\bf  r} + (h+\lambda u){\bf e}_z) \left[ 1-\left(1+\frac{u}{2}\right)\,e^{{-u}/{2}}
\right]
\,\d u, \nonumber
\end{eqnarray}
which leads to asymptotic formulae for the image system using an expansion of the terms in Eq.~\eqref{imagepara_scaled}. In the limit $\kn \ll 1$, the image system is found to be given  by
\begin{equation}
\w^\parallel(\r,\lambda)\approx 2F[-\G^S_x+h\G^{SD}_{z;x}-h^2\G^D_x](\bar {\bf  r})
+4\lambda F [\G^{SD}_{x;z}-h\G^{SQ}_{z;zx}+h^2\G^{Q}_{xz}](\bar {\bf  r}),
\end{equation}
which is Blake's solution for a no-slip surface plus an $\oo{\kn}$ correction. 
Note that the rotlet dipole in Eq.~\eqref{imagepara_scaled}, $\G^{RD}_{z;y}$, will only appear in asymptotic formulae for the image system at order $\oo{\kn^2}$.

\subsection{Image system for a point source}

We finally consider in this section the image system for a point
source, which we will denote by $^{(.)}$. In this case, the
velocity field and its first image are given by
\begin{subeqnarray}
\U\. & = & \frac{Q}{8\pi\mu}\frac{\r}{r^3}, \\
\V\. & = & \frac{Q}{8\pi\mu}\frac{\bar \r}{\bar r^3},
\end{subeqnarray}
where $4\pi Q$ can be interpreted as the source flow rate. Here
again, we decompose the velocity field $\u\.=\U\.+\V\.+\w\.$ and
need to solve for $\w\.$. The boundary conditions for $\w\.$ are
exactly proportional to those for $\w^\perp$ (Eq.~\ref{samebc}); correspondingly, the two solutions are proportional as well, giving
\begin{equation}\label{}
\w\.=-\frac{2Q}{Fh}\w^\perp = -\frac{2Q}{\lambda} \int_0^\infty
e^{{-s}/{2\lambda}} [(h+s)\,\G^D_z-\G^{SD}_{z;z}] ({\bf  r} + (h+s){\bf e}_z)\d s.
\end{equation}

\subsection{Higher-order singularities}

Finally, we note that higher-order singularities may be derived from the Stokeslet and point source image systems presented above, by taking derivatives with respect to the singularity location.  Subtleties do exist, however, so care should be taken \cite{blake74}.  Derivatives along the plane of the wall can be computed in a straightforward fashion, but derivatives perpendicular to the wall are more subtle, because the amplitude of the image singularities depends on the distance from the wall.  In general, the correct image system will always be obtained if the derivative is calculated by taking the limit of two such opposing singularities (since each solution obeys the correct boundary condition on the wall in the first place).   That is, if ${\bf u}^{{\rm sin}}(h)$ is the fundamental singularity located a distance $h$ from the wall, and ${\bf u}^i(h)$ is the proper image, then the image system for a perpendicular dipole of  ${\bf u}^{{\rm sin}}(h)$ can be found by taking the limit
\be
{\bf u}^D=\lim_{\epsilon \rightarrow 0} \left(\frac{ {\bf u}^{{\rm sin}}(h)+{\bf u}^i(h)- {\bf u}^{{\rm sin}}(h-\epsilon) - {\bf u}^i(h-\epsilon)}{\epsilon}\right).
\ee

\section{Influence of slip on Brownian motion}
\label{Influence}
\subsection{Diffusion of a spherical particle}

We now turn to a specific application of the above calculation:  the diffusivity ${\bf D}$ of a solid spherical particle of radius $a$ near a partial-slip wall.  Using the Stokes-Einstein relation, the diffusivity is directly proportional to the particle mobility ${\bf b}$ via ${\bf D}=k_BT{\bf b}$; thus the (deterministic) calculation of mobility yields the diffusivity. 

Particle mobilities are defined as the velocity response to a force ${\bf F}$ acting on the particle. In the absence of solid boundaries, and if the no-slip boundary condition is satisfied on the particle surface, the velocity field established around the particle is given by
\begin{equation}\label{Stokeslet}
{\bf u}=\frac{1}{8\pi \mu}\left(\frac{{\bf
1}}{r}+\frac{{\bf r}{\bf r}}{r^3} \right)\cdot{\bf F} + \frac{a^2}{24\pi\mu}
\left(\frac{{\bf 1}}{r^3}-\frac{3{\bf r}{\bf r}}{r^5}
\right)\cdot{\bf F},
\end{equation}
leading to the (isotropic) Stokes mobility 
\be
{\bf b}_0=\frac{{\bf 1}}{6\pi\mu a}.
\label{eq:stokesmobility}
\ee
The first term in Eq.~\eqref{Stokeslet}, which decays like $1/r$, corresponds to a Stokeslet, whereas the second term (source dipole) is necessary to satisfy the no-slip boundary condition on the surface of the particle, and decays like $1/r^3$.

The presence of a nearby surface (a distance $h$ from the particle) modifies the flow field around the particle, and hence its mobility, which is now an anisotropic tensor ${\bf b}$.  To account for such effects, one can employ an approximate and iterative technique known as the method of reflections \cite{happel83}.  When the particle is `far' from the wall ($a\ll h$), it sets up a flow field that is locally appears like that around a particle in an infinite fluid.  This flow field, however, violates the boundary conditions at the wall, and so an additional field (`reflection') is introduced to correct the boundary conditions at the wall.  This first reflection, however, violates the boundary conditions at the particle surface, necessitating a second reflection, and so on. 

Thus in our approximation, the particle travels through its local fluid environment with Stokes mobility \eqref{eq:stokesmobility}, and is advected by the image flow $\u^{w}$ via Faxen's law
\be
\u^{{\rm adv}} = \u^w +\frac{a^2}{6}\nabla^2\u^w.
\label{eq:faxen}
\ee
Thus the leading order correction (in $a$) to the mobility is given by the wall-induced flow $\u^{{\rm w}}$ evaluated at the particle location.  Furthermore, to obtain the $\oo{a}$ component of $\u^{{\rm w}}$, only the image system for the Stokeslet in \eqref{Stokeslet} is required.  Errors to this approach are of order $a^3/h^3$, since the source dipole flow in \eqref{Stokeslet} and the Laplacian in \eqref{eq:faxen} are smaller by $a^2$.   Note also that, since only the image system for the Stokeslet is required for the leading-order wall correction to the mobility, the result is insensitive to the boundary conditions ({\sl i.e.} size, shape, or slip) on the particle itself.

\subsection{The effect of slip upon single-particle diffusivity}

\begin{figure}[t]
\centerline{
\includegraphics[width=3in]{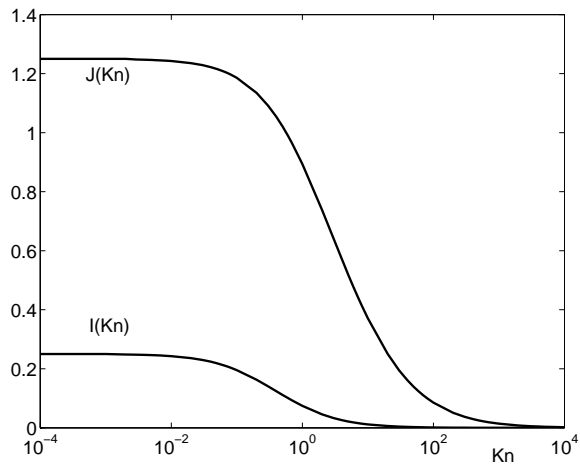}
}
\caption{Variations of $\I$ and $\J$ with the Knudsen number $\kn$.} 
\label{fig:plots}
\end{figure}

The components of the velocities at the position of the Stokeslets and in the same direction as the applied force, denoted
generically $(u_1,u_2,u_3)$, are given by
\begin{eqnarray}
u_3^\perp&=&\frac{F}{6\pi\mu
a}+V^\perp_3(0,0,h)+w^\perp_3(0,0,h),\label{mob_perp}
\\
u^\parallel_1&=&\frac{F}{6\pi\mu
a}+V^\parallel_1(0,0,h)+w^\parallel_1(0,0,h)\label{mob_parallel}.
\end{eqnarray}
The first terms ($\U^\perp,\U^\parallel$) represent the bulk (Stokes) mobility; the second terms ($\V^\perp,\V^\parallel$) represent advection with the flow field established by the primary image Stokeslets, and the third terms ($\w^\perp,\w^\parallel$) reflect advection with the higher-order image field, as detailed in Appendices \ref{solve_perpendicular} and \ref{solve_parallel}.  The flow velocities from the primary image Stokeslet, evaluated at the particle position, are
\begin{equation}
V^\perp_3(0,0,h)=-\frac{F}{8\pi \mu h}, \,\,\,
V^\parallel_1(0,0,h)=\frac{F}{16\pi \mu h}, 
\end{equation}
and inverting the Fourier transforms from Appendices
\ref{solve_perpendicular} and \ref{solve_parallel} reveals the contribution from higher-order singularities to be
\begin{subeqnarray}
w^\perp_3(0,0,h) & = & -\frac{F}{4\pi\mu h}\,\I
\left(\kn\right), \\
w^\parallel_1(0,0,h) & = & -\frac{F}{8\pi\mu
h}\,\J\left(\kn\right)
\end{subeqnarray}
where the functions $\I$ and $\J$ are defined by
\begin{subeqnarray}
\I(\kn) & = &  \int_0^{\infty}
\frac{x^2}{1+2x\kn}\,e^{-2x}\d x = \frac{\kn (\kn -1)+e^{1/\kn}\Gamma(0,1/\kn)}{8\kn^3}, \\
\J(\kn) & = & \int_0^{\infty}  \frac{(1+\kn
r)(1-r)^2+2(1+2\kn r)}{(1+\kn r)(1+2\kn r)} \,e^{-2r}\d r\\
&=& \frac{-\kn (3\kn+1)+e^{1/\kn}(1+2\kn)^2\Gamma(0,1/\kn)}{8\kn^3} + \frac{e^{2/\kn}\Gamma(0,2/\kn)}{\kn}, \nonumber
\end{subeqnarray}
where $\Gamma$ is the incomplete Gamma function \cite{abramowitz72}
\begin{equation}
\Gamma(a,x)=\int_x^\infty t^{a-1}e^{-t}\,\d t.
\end{equation}
The results of Eqs.~\eqref{mob_perp} and \eqref{mob_parallel} yield the desired relation between wall slip and nearby colloidal diffusivities, given to leading order in $a/h$ by
\begin{subeqnarray}
\frac{D^\perp}{D_0}=\frac{b^\perp}{b_0}&=&1-\frac{3a}{4h}\left(1+2{\cal
I}\left(\kn\right)\right) + {\cal O}\left(
\frac{a^3}{h^3} \right) \slabel{change_perp}, \\
\frac{D^\parallel}{D_0}=\frac{b^\|}{b_0}&=&1+\frac{3a}{8h}\left(1-2{\cal
J}\left(\kn\right) \right) + {\cal O}\left(
\frac{a^3}{h^3} \right) \cdot\slabel{change_parallel}
\end{subeqnarray}
The functions $\I$ and $\J$ are illustrated in Fig.~\ref{fig:plots}.  It is significant to note that appreciable variation in parallel diffusivity ($\J$) occurs over about three decades in $\kn$, giving a fairly wide range of experimental conditions under which one might hope to measure $\lambda$.

\subsection{Asymptotic limits}
\label{asymptot}

When the slip length is small ($\lambda\ll h$), we can use a Taylor
expansion of Eqs.~\eqref{change_perp} and
\eqref{change_parallel} and obtain the asymptotic results
\begin{subeqnarray}\label{change}
\frac{D^\perp}{D_0} & = & 1-\frac{9a}{8h}\left(1-\kn
+ {\cal O}\left( \kn^2 \right) \right) + {\cal
O}\left( \frac{a^3}{h^3} \right), \\
\frac{D^\parallel}{D_0} & = &
1-\frac{9a}{16h}\left(1-\kn + {\cal O}\left(
\kn^2 \right)  \right) + {\cal O}\left(
\frac{a^3}{h^3} \right).
\end{subeqnarray}
The results for a no-slip surface, Eq.~\eqref{noslipdiff}, are therefore
recovered as the slip length vanishes.  The `slightly slipping' result can also be
obtained simply from the no-slip solution, using a reciprocal theorem for Stokes flow,
as shown in Appendix  \ref{recip}.

When the slip length is large $\lambda \gg h$, on the other hand, 
Eqs.~\eqref{change_perp} and \eqref{change_parallel} can be approximated by
\begin{subeqnarray}
\frac{D^{\perp}}{D_0} & = &
1-\frac{3a}{4h}\left(1+\frac{1}{4\kn} + {\cal O}\left(
\frac{1}{\kn^2} \right)\right) + {\cal O}\left(
\frac{a^3}{h^3} \right),
\\
\frac{D^{\parallel}}{D_0} & = & 1+\frac{3a}{8h}\left(1+
\frac{5}{\kn} \ln\left(  \frac{1}{\kn}\right) + {\cal
O}\left( \frac{1}{\kn} \right)   \right) + {\cal O}\left(
\frac{a^3}{h^3} \right).
\end{subeqnarray}
The results for a no-shear surface, Eq.~\eqref{slipdiff}, are
recovered when the slip length diverges, albeit slowly (note the logarithmic
dependence in the parallel case).

\section{Coupled diffusion of two particles near a partial slip surface}
\label{correlated}

We extend in this section the idea proposed in \S\ref{Influence}  to the coupled diffusivity of two colloids. We consider two spherical particles, radius $a$, located at distance $d$ from each other along the $x$-axis and at a distance $h$ above the slipping surface. 

The diffusivity tensor, ${\bf D}$, for a general N-particle system is given by \cite{batchelor76}
\begin{equation}
{\bf D}=k_B T\, {\bf b}
\end{equation}
where ${\bf b}$ is the N-particle mobility tensor. In the case of two particles, these mobilities are the tensors  relating the forces, ${\bf F}_1$ and ${\bf F}_2$, acting on each particle and their velocities, ${\bf u}_1 $ and ${\bf u}_2 $, as
\begin{subeqnarray}
{\bf u}_1 & = & {\bf b}_{11}\cdot {\bf F}_1 + {\bf b}_{12}\cdot {\bf F}_2, \\
{\bf u}_2 & = & {\bf b}_{21}\cdot {\bf F}_1 + {\bf b}_{22}\cdot {\bf F}_2. 
\end{subeqnarray}

The values of the tensors ${\bf b}_{11}={\bf b}_{22}$ reflect the self-diffusion of each particle, and the influence of slip on their values was presented in \S\ref{Influence}. We calculate below the influence of slip on the coupling tensors, ${\bf b}_{12}^T={\bf b}_{21}$. Note that unlike ${\bf b}_{11}$ and ${\bf b}_{22}$, the coupling tensors ${\bf b}_{12}$ and ${\bf b}_{21}$ do not depend on the particle size $a$, but rather on the value of the new length scale, $d$.

Since there is  symmetry between particles 1 and 2, we have  ${\bf b}_{12}^T={\bf b}_{21}$, and it is therefore sufficient to  calculate  only ${\bf b}_{21}$. Furthermore, since the particles are aligned along the $x$-direction, the tensor has only five non-zero entries,
\begin{equation}
{\bf b}_{21}=\left(
\begin{array}{ccc}
b_{x_2x_1}& 0 & b_{x_2z_1}\\
0 &b_{y_2y_1}&0\\
b_{z_2x_1}& 0 & b_{z_2z_1}
\end{array}\right).
\end{equation}
By symmetry, $b_{x_2z_1}=-b_{z_2x_1}$, leaving just four independent components in the coupling mobility/diffusivity.

We adopt the same method as in \S\ref{Influence} and consider the limit $a\ll h$, so that we replace the full velocity field around the particle by a Stokeslet.  In the case of no-slip, the coupled mobilities (influence of the point force and its image system) are  given by 
\begin{equation}
{\bf b}_{21}(\kn=0)=\frac{1}{8\pi \mu d}\left(
\begin{array}{ccc}
2\left(\displaystyle 1-\frac{1+\xi+\frac{3}{4}\xi^2}{(1+\xi)^{5/2}}\right) & 0 & \displaystyle \frac{-\frac{3}{2}\xi^{3/2}}{(1+\xi)^{5/2}}\\
0 &\displaystyle 1-\frac{1+\frac{3}{2}\xi}{(1+\xi)^{3/2}}&0\\
 \displaystyle \frac{\frac{3}{2}\xi^{3/2}}{(1+\xi)^{5/2}}& 0 & \displaystyle 1-\frac{1+\frac{5}{2}\xi + 3\xi^2 }{(1+\xi)^{5/2}}
\end{array}\right)
\end{equation}
whereas when the slip length is infinite, they are given by
\begin{equation}\label{coupled_perfect}
{\bf b}_{21}(\kn=\infty)=\frac{1}{8\pi \mu d}\left(
\begin{array}{ccc}
2\left(\displaystyle 1+\frac{1+\frac{1}{2}\xi}{(1+\xi)^{3/2}}\right)& 0 &\displaystyle -\frac{\xi^{1/2}}{(1+\xi)^{3/2}}\\
0 &\displaystyle 1+\frac{1}{(1+\xi)^{1/2}}&0\\
\displaystyle \frac{\xi^{1/2}}{(1+\xi)^{3/2}} & 0 & \displaystyle 1-\frac{1+2\xi}{(1+\xi)^{3/2}} 
\end{array}\right)
\end{equation}
with 
\be
\xi=\frac{4h^2}{d^2}.
\ee
These results are correct to order $\oo{a^3/h^3}$ and $\oo{a^3/d^3}$. The qualitative difference between the 
no-slip and perfect slip formulae concern their spatial decay in the limit $\xi \to 0$ ($ h \ll d$). In that case,  all components of the mobility tensor decay faster in the case of no-slip than in the case of perfect slip. 

We present below the calculation for  $b_{z_2z_1}$, as it is the configuration where calculations are the easiest. Calculations for other components of the mobility tensor are derived in Appendix~\ref{mobility_tensor}, with results summarized below as well. 

In order to evaluate  $b_{z_2z_1}$, we consider a unit vertical force applied to particle 1, and determine the vertical velocity at the position of particle 2.  Two factors contribute to the value of $b_{z_2z_1}$:   the direct influence of the Stokeslet flow field from particle 1 and the influence of the image system for this Stokeslet below the slipping surface. Given the decomposition assumed in Eq.~\eqref{decomp}, the mobility is given by 
\begin{equation}
b_{z_2z_1}(\kn)=b_{z_2z_1}(\kn=\infty) + w_3^\perp(d{\bf e}_x+h{\bf e}_z,\lambda).
\end{equation}
Using Eq.~\eqref{coupled_perfect} and and evaluating the integral in Eq.~\eqref{interpretation_perp} leads to 
\begin{equation}\label{bzz}
b_{z_2z_1}(\kn)=\frac{1}{8\pi \mu d} \left[
1-\frac{1+2\xi}{(1+\xi)^{3/2}} 
+ \frac{\xi}{2 }\left(\frac{1}{\kn}\int_0^\infty e^{-u/
\kn}\frac{1-2\xi(1+u)^2}{(1+\xi(1+u)^2)^{5/2}}
\,\d u.\right)\right]
\end{equation}
For a fixed value of $\kn$, the asymptotic behavior of the integral in parenthesis in Eq.~\eqref{bzz} in the limit $\xi\to 0$ is given by
\begin{equation}\label{integral_asymptotics}
\frac{1}{\kn}\int_0^\infty e^{-u/
\kn}\frac{1-2\xi(1+u)^2}{(1+\xi(1+u)^2)^{5/2}}
\,\d u = 1 -\frac{9}{2} (1 + 2\kn + 2\kn^2)\,\xi +{\cal O}(\xi^2).
\end{equation}
Consequently, for any value of the slip length, there exists a range of $\xi$, 
\begin{equation}
\xi \ll \min\left(1,\frac{1}{\kn^2}\right),
\end{equation}
which corresponds to the far-field limit
\begin{equation}
d^2 \gg \max (h^2,\lambda^2),
\end{equation}
for which the integral in Eq.~\eqref{integral_asymptotics} goes to one asymptotically. It follows that, evaluating Eq.\eqref{bzz}, the mobility always decays asymptotically  as ${\cal O}\left(\xi^2/{\mu d} \right)\sim {1}/{d^5} $; this is the same power law as the no-slip case. The case of perfect slip is therefore a singular limit: as $\kn\to\infty$, the range of $\xi$ for which this asymptotic behavior is valid,  $\xi \ll 1/\kn^2$, shrinks to zero, resulting in an asymptotic behavior of the perfect-slip diffusivities  qualitatively different from that of any other partially slipping surface.  We can then use these results to obtain, for a given slip length, the asymptotic behavior of the mobility as $\xi\to 0$. Substituting the result of Eq.~\eqref{integral_asymptotics} in Eq.~\eqref{bzz}, we obtain
\begin{equation}\label{slip_asymptotics}
b_{z_2z_1}(\kn)=- \frac{9\xi^2}{64\pi \mu d} (1+4\kn+4\kn^2)\,\left[1+{\cal O}(\xi)\right].
\end{equation}
As is obvious in Eq.~\eqref{slip_asymptotics}, the numerator is a function of $\kn$, and as consequence, a measure of the behavior of the spatial decay of the coupled diffusivities of the two particles allows, in principle,  to infer the value of the slip length.

The other components of the mobility tensor ${\bf b}_{21}$ are calculated in Appendix~\ref{mobility_tensor} for small values of $\xi$. They are given by 
\begin{eqnarray}
b_{x_2x_1}(\kn) & = & \frac{3\xi}{8\pi \mu d}\left(1+2\kn + \kn^2\right)\,[1+\oo{\xi}], \label{xx}\\
b_{y_2y_1}(\kn) & = & \frac{3\xi^2}{64\pi \mu d}\left(1+4\kn + 12\kn^2+16\kn^3+8\kn^4\right)\,[1+\oo{\xi}],\label{yy}\\
b_{x_2z_1}(\kn) & = &-\frac{3\xi^{3/2}}{16\pi \mu d} \left(1+3\kn+2\kn^2 \right)\,[1+\oo{\xi}]. \label{xz}
\end{eqnarray}
and similarly, their measurement would allow an estimation of the slip length on the surface; Eqs.~\eqref{xx} and \eqref{yy} are valid when $d \gg \max(h,\lambda)$ and Eq.~\eqref{xz} when $d^2 \gg \max(h^2,\lambda^2)$. Note that the largest leading-order influence of a non-zero slip length are obtained for the components $b_{z_2z_1}$ and $b_{y_2y_1}$ of the coupled matrix (behavior $\approx 1+4\kn$ for small $\kn$), although these components decay most quickly with $\xi$.  The most slowly-decaying coupling mobility is $b_{x_2x_1}$.

\section{Discussion and Conclusions}

Having presented the fundamental singularities for Stokes flow near a partial-slip planar wall and explored the consequences for colloidal diffusion, we now turn to examine issues relevant to experimental studies of slip.

Before we begin, it is significant to note that there are several length scales inherent in the systems we have been considering:  the colloidal radius $a$, distance from the wall $h$, slip length $\lambda$, and (for multi-particle systems) the distance $d$ between the particles.  The theory presented above concerns the fluidic response to a point force, and thus the results we presented are valid for particles that are `far' from the wall, so that $a \ll h$. (Note, however, that excellent agreement between theory and experiment was obtained for the coupled colloidal diffusion even for systems with $a/h\sim 1/3$ \cite{dufresne00}.)  This is significant because the transition between `no-slip' and `perfect slip' occurs, not surprisingly, around $\kn \sim \oo{1}$, or $h \sim \lambda$.  Thus to experimentally observe the transition between the two slip regimes (and thus measure $\lambda$ convincingly), a tracer is required that is of the same order as the slip length itself. 

Different slip lengths could be probed with different experimental techniques.  Several experiments 
\cite{schnell56,boehnke99,zhu01,tretheway02,lumma03} have reported slip lengths of order microns, which would allow the use of micron-sized colloids.  Optical tweezers can trap colloids of this size, and thus allow their repeatable and precise three-dimensional placement.  Repeatedly trapping and releasing single or multiple colloids has proven an excellent method for measuring spatially-varying single- or multi-particle diffusivities \cite{crocker97,dufresne00,dufresne01}, and would thus be naturally adaptable to probe the slip properties of walls as described above.  Such studies have been developed using video microscopy, which most typically would measure motion parallel to the wall, and project out perpendicular motion.
 Again, this requires a probe that is large enough to be trapped by optical tweezers, which places a lower limit on the slip length that could practically be measured, although a null measurement using tweezers would be useful in putting an upper bound on the slip length of a surface.
 
Other slip experiments \cite{churaev84,kiseleva99,pit00,baudry01,craig01b,cheng02,cottinbizonne02,sun02,zhu02d,bonaccurso03,cheikh03,choi03,lumma03,neto03,cho04,henry04,cottin-bizonne05,joseph05} 
report shorter ($10-100$~nm) slip lengths, which would require smaller ($10-100$~nm) tracers.  For visualization, such tracers should presumably be fluorescent, such as quantum dots (see, e.g., \cite{lumma03})   These small colloids are more difficult, if not impossible, to hold with optical tweezers, which rules out their precise manipulation and placement.  Instead, techniques such as total internal reflection microscopy allow accurate three-dimensional measurements of their positions, and could thus be used to probe surfaces with $10-100$~nm slip lengths.  Such techniques have the additional advantage of precise, three-dimensional measurements, and could thus be used to probe both the perpendicular and parallel diffusivities.  Additional issues also arise with such small tracers:  obviously, the diffusive motion is significantly higher, so correspondingly particle motion must be resolved on faster time scales.  Furthermore, colloid/wall interactions (electrostatic, van der Waals, and so on \cite{russel89}) can become significant at these shorter length scales, and must be treated properly in data analysis.

We now discuss issues specific to single-particle diffusivities.  From Eqs.~\eqref{noslipdiff} and \eqref{slipdiff}, it is evident that variations in $\kn$ give rise to a change in single-particle diffusivities of at most 
\begin{equation}
\frac{\Delta D^\perp}{D_0}=\frac{3a}{8h}, \quad \frac{\Delta
D^\parallel}{D_0}=\frac{15a}{16h},
\end{equation}
to leading order in $a/h$. The effect on the diffusivity parallel to the wall is larger than for the perpendicular diffusivity, and furthermore changes sign:  parallel diffusivity is hampered by a no-slip wall, but enhanced by a significantly slipping wall, with a crossover occuring at $\kn \approx 5.45$, where $\J (\kn) = 1/2$.  Furthermore, from Fig. \ref{fig:plots}, the parallel diffusivity changes over about three decades in $\kn$, providing a further advantage to the parallel mode of measurement.  In principle, significantly smaller particles ($a \ll h \sim \lambda$) could be used in such measurements; however, the correction to the self-diffusivity is of order $a/h$, and thus the smaller the probe, the smaller the effect to be measured.

\begin{figure}[t]
\centerline{
\includegraphics[width=.6\textwidth]{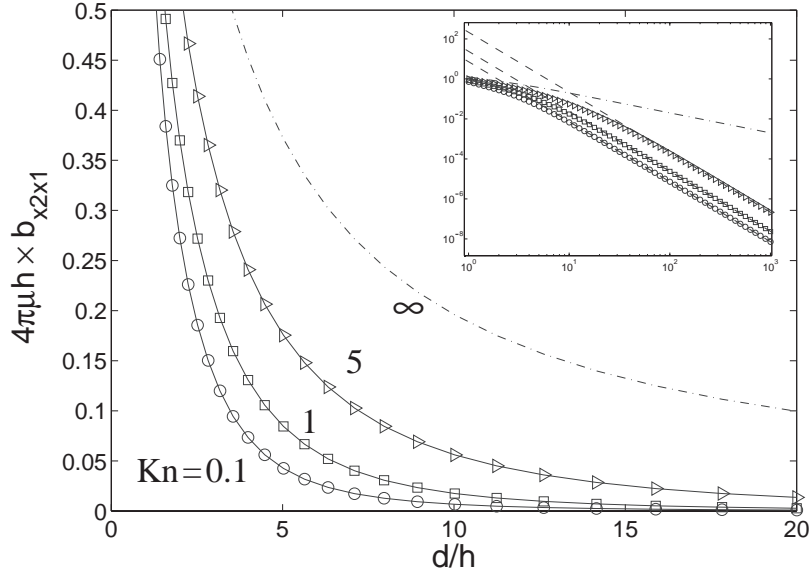}}
\caption{Variation of the coupled mobility $b_{x_2x_1}$ (non-dimensionalized by $1/4\pi \mu h$) with $d/h$, for four values of the Knudsen number: $\kn=0.1$ (circles and solid line),  $\kn=1$ (squares and solid line), $\kn=5$ (triangles and solid line) and $\kn=\infty$ (dashed-dotted line).  Insert: same data but in log-log scale, with included the asymptotic behaviors for each value of $\kn<\infty$ as given in Eq.~\eqref{xx} (dashed lines).} 
\label{bxx_plot}
\end{figure}

Multi-particle diffusion, on the other hand, affords a greater variety of measurements.  In \cite{crocker97} and \cite{dufresne00}, pair diffusivities were measured in terms of center-of-mass and relative motion variables, wherein corrections due to hydrodynamic interactions are smaller than the bulk `Stokes' diffusivity by factors of order $a$.  Rather than measuring these particular modes, we suggest simply measuring the cross-correlation between two probes:
\be
\frac{1}{2} \frac{d}{dt} \langle \Delta \alpha_1\Delta \beta _2\rangle = k_B T b_{\alpha_1\beta_2},
\ee
where $\alpha$ and $\beta$ can take the values $\{x,y,z\}$ and $b$ represents the coupling mobility.  Notably, the coupling mobilities depend on neither the shape nor the size of the colloids themselves.  (This feature plays a significant role in so-called two-point microrheology \cite{crocker00}.)  From Eqs.~\eqref{slip_asymptotics}-\eqref{xz}, one can see that the largest effect of slip upon cross-correlated diffusive  motion occurs for the $zz$ and $yy$ modes, although the $xx$ has the slowest spatial decay and may be easiest to measure (see Fig.~\ref{bxx_plot}).

Another attractive feature of the proposed experimental system is the ease of performing multiple experiments within the same experimental cell.  Surfaces with patterned properties could be used to probe surfaces with different putative slip lengths, and differential measurements could be used to remove the uncertainties associated with cell-to-cell variability.  

Although the calculations presented here are valid only for particles `far' from the wall, the experiments proposed here need not be performed in this limit.  In fact, the results we present here could assist in boundary-integral studies to obtain the various mobilities/diffusivities for systems with colloidal spheres `near' partially-slipping walls.

Finally, although the idea proposed here is concerned primarily with {\it passive} microrheology, the change in the particles mobilities could in theory also be measured using {\it active} microrheology, {\it i.e.}, measuring the direct relationship between particle motion and a known applied force, whether that force were applied to the particle itself (in which case the self-mobility would be measured), or to an adjacent particle (in which case the coupling mobility would be measured). 

\section*{Acknowledgements}
We thank  L. Bocquet and H. Chen for useful discussions.  EL gratefully acknowledges support of the Office of Naval Research (Grant \# N00014-03-1-0376) and the Harvard MRSEC.  TMS gratefully acknowledges the support of the Lee A. Dubridge Prize Postdoctoral Fellowship and the NSF Mathematical Sciences Postdoctoral Fellowship.
\appendix

\section{General solution of the Stokes equations near a planar boundary}
\label{solution}

We present in this section the general solution of Stokes equation
near a solid boundary, using the notations introduced in \S
\ref{setup}, for a general velocity field $\u$. We introduce the
Fourier transform in the directions parallel to the plane, 
for each of the velocity components
\begin{equation}
\tilde u_j(k_1,k_2,z)={\cal F}[u_j]= \frac{1}{2\pi}\int\int
u_j(x,y,z)\,e^{ik_1x+ik_2y}\,\d x\,\d y, \quad (j=1,2,3)
\end{equation}
and its inverse
\begin{equation}
u_j(x,y,z)={\cal F}^{-1}[\tilde u_j]= \frac{1}{2\pi}\int\int
\tilde u_j(k_1,k_2,z)\,e^{-ik_1x-ik_2y}\,\d k_1\,\d k_2.
\end{equation}
The general solution to Eq.~\eqref{stokes} is given by
\begin{subeqnarray}\label{general}
\tilde{u}_\alpha & = & \left(\frac{ h}{8\pi \mu}\right) \left\{A_\alpha+ik_\alpha Bz \right\}e^{-kz},\quad \alpha=\{1,2\}\\
\tilde{u}_3 & = & \left(\frac{ h}{8\pi \mu}\right) \left\{A_3+ k
Bz\right\}e^{-kz}
\end{subeqnarray}
where $k=(k_1^2+k_2^2)^{1/2}$ and $\alpha$ can take the values $\{1,2\}$, and where we have taken the magnitude of the force to be $F=1$ to simplify the notations (as the equations are linear with respect to $F$, this can be done without loss of generality). The four dimensionless constants ($A_1,A_2,A_3,B$) are linked through the
continuity equation, giving
\begin{equation}
i(k_1A_1+k_2A_2)=k(B-A_3).
\end{equation}
The remaining three constants are found by applying the slip
boundary conditions, Eq.~\eqref{newbound}, on the surface, which determines the velocity
field uniquely.

We note for future use that the Fourier transform of the velocity field \eqref{general} can be divided into two components:
\begin{equation}
\tilde{{\bf u}} = \tilde{{\bf u}}_a + z \tilde{{\bf u}}_b,
\end{equation} 
giving corresponding real-space velocity fields
\begin{equation}
{\bf u} = {\bf u}_a + z {\bf u}_b,
\end{equation} 
where ${\bf u}_a = {\cal F}^{-1}[\tilde {\bf u}_a]$, and ${\bf u}_b = {\cal F}^{-1}[\tilde {\bf u}_b]$.

%%%%%%%%%%%%%%%%%% PERP %%%%%%%%%%%%%%
\section{Solution for perpendicular Stokeslet}
\label{solve_perpendicular}

The boundary conditions, Eq.~\eqref{newbound}, for the flow $\w^\perp$ due to a Stokeslet perpendicular to the surface 
are thus given by the no-flux condition, $w_3^\perp=0$, and the slip condition
\begin{equation}\label{samebc}
\left[w_\alpha^\perp-\lambda \frac{ \partial w_\alpha^\perp}{\partial z}\right](x,y,0) =-2U_\alpha^\perp=\frac{ h}{4\pi
\mu}\left(\frac{x}{r_h^3},\frac{y}{r_h^3}\right),
\end{equation}
where $r_h^2=x^2+y^2+h^2$ and $\alpha=\{1,2\}$. In Fourier space, these conditions
become $\t w_3^\perp=0$ and
\begin{equation}\label{perpj}
\left[\t w_\alpha^\perp -\lambda \frac{ \partial \t w_\alpha^\perp}{\partial z}\right](k_x,k_y,0)= i \frac{ h}{4\pi \mu}  \frac{k_\alpha}{k}e^{-k h}.
\end{equation}
Using the general formalism of Eq.~\eqref{general} together with the boundary conditions Eq.~\eqref{perpj},  we obtain
\begin{subeqnarray}\label{As_perp}
A_1^\perp&=&\frac{2ik_1}{k(1+2\lambda k)}\,e^{-k h},\\
A_2^\perp&=&\frac{2ik_2}{k(1+2\lambda k)}\,e^{-k h},\\
A_3^\perp&=&0,\\
B^\perp&=&\frac{-2}{1+2\lambda k}\,e^{-kh}.
\end{subeqnarray}
Writing these constants as
\begin{subeqnarray}
A^\perp_\alpha(\k,z,\lambda)&=&\frac{A^\perp_\alpha(\k,z,0)}{1+2\lambda k},\\
B^\perp(\k,z,\lambda)&=&\frac{B^\perp(\k,z,0)}{1+2\lambda k},
\end{subeqnarray}
where $A^\perp_j(\k,z,0)$ and $B^\perp(\k,z,0)$ are the Fourier coefficients for Blake's no-slip case, allows us to express the Fourier components of the velocity field as
\begin{equation}
 \tilde {\bf w}^\perp(\r,\lambda)={ \tilde{\bf w}}_a^\perp(\r,\lambda) + z \,{\tilde {\bf
w}}_b^\perp(\r,\lambda),
\label{B5}
\end{equation}
where 
\begin{subeqnarray}
\left(1-2\lambda \frac{\p}{\p z} \right){\tilde {\bf
w}}_a^\perp(\r,\lambda)& = & {\tilde {\bf w}}_a^\perp(\r,0),
\\\slabel{int1_parallel}
\left(1-2\lambda \frac{\p}{\p z} \right){{ \tilde{\bf
w}}}_b^\perp(\r,\lambda)& = &{{ \tilde {\bf
w}}}_b^\perp(\r,0)\cdot\slabel{int2_parallel}
\end{subeqnarray}
Differentiating Eq.~\eqref{B5}  and inverting the Fourier transforms leads to
\begin{equation}\label{parallel_tointegrate}
\left(1-2\lambda \frac{\p}{\p z} \right) {\bf w}^\perp(\r,\lambda)
=
 {\bf w}^\perp(\r,0)
 -2\lambda { \,{ {\bf
w}}_b^\perp}(\r,\lambda).
\end{equation}
The solution ${\bf w}^\perp(\r,0)$ is such that $\V^\perp(\r)+{\bf w}^\perp(\r,0)$
is Blake's solution for a no-slip surface, that is,
${\bf w}^\perp(\r,0)=2h^2\G_z^D-2h\G_{z;z}^{SD}$, 
where $\G^D_z$ represents the source dipole in the $z$-direction,
\be
\G^D_z({\bf r})=-\pd{}{z} \left(\frac{1}{8\pi\mu}\frac{{\bf r}}{{\bf }r^3} \right) 
\ee
and $\G^{SD}_{z;z}$ represents a $z$-dipole of z-Stokeslets \cite{blake71,pozrikidis92},
\be
\G^{SD}_{z;z}({\bf r})=-\pd{}{z}\left({\bf {e}}_z \cdot {\bf G}^S{\bf (r)}\right).
\ee
Furthermore, if we recognize that
\begin{equation}
 {\bf
w}_b^\perp(\r,0)=-2h\G_z^D(\r),
\end{equation}
we can integrate each term in Eq.~\eqref{parallel_tointegrate} using Eq.~\eqref{int2_parallel} and integration by parts to obtain
\begin{equation}\label{interpretation_perp2}
\w^\perp(\r,\lambda)=\frac{h}{ \lambda} \int_0^\infty
e^{{-s}/{2\lambda}} [(h+s)\,\G^D_z-\G^{SD}_{z;z}]({\bf  r} + (h+s){\bf e}_z)\d s,
\end{equation}

Thus the complete image system for a Stokeslet oriented perpendicular to a partial-slip wall is given by  $\u^\perp=\U^\perp+\V^\perp+\w^\perp$, where $\U^\perp$ and
$\V^\perp$ are given by Eqs.~\eqref{Uperp} and \eqref{Vperp}, and where
$\w^\perp(\r,\lambda)$ is given by \eqref{interpretation_perp2}.
Notably, this is the same image system as in Blake's no-slip solution; in the partial-slip case, however, the image singularities are distributed along a line in the $-{\bf e}_z$ direction, with a magnitude that decays exponentially over $2\lambda$.

%%%%%%%%%%%%%%%%%% PARA %%%%%%%%%%%%%%
\section{Solution for parallel Stokeslet}
\label{solve_parallel}

The image system for a Stokeslet oriented parallel to a partial-slip wall is more complicated than for the perpendicular Stokeslet, but conceptually similar. 
In this case,  Eq.~\eqref{newbound} for
$\w^\parallel$ become $w_3^\parallel=0$ and
\begin{equation}
\left[w_\alpha^\parallel-\lambda \frac{ \partial w_\alpha^\parallel}{\partial z}\right](x,y,0)= -2U_\alpha^\parallel =-\frac{1}{8\pi
\mu}\left(\frac{2}{r_h}+\frac{2x^2}{r_h^3},\frac{2xy}{r_h^3}\right),
\end{equation}
with $r_h^2=x^2+y^2+h^2$ and $\alpha=\{1,2\}$. In Fourier space, and using
Eq.~\eqref{fourier2}, these conditions become $\t w_3^\parallel=0$ and
\begin{equation}\label{parallelj}
\left[\t w_\alpha^\parallel -\lambda \frac{ \partial \t
w_\alpha^\parallel}{\partial z}\right] (k_x,k_y,0)=-\frac{1}{8\pi \mu}\left(
2\frac{k_1^2(1-hk)+2k_2^2}{k^3},\frac{-2k_1k_2(1+hk)}{k^3}
\right){e^{-kh}}.
\end{equation}

Here again we can now solve for $\t\w^\parallel$, using the
general formalism given by Eq.~\eqref{general}, gives
\begin{subeqnarray}\label{As_parallel}
A_1^\parallel&=&-\frac{2}{hk^3}\left[\frac{k_1^2(1-hk)(1+\lambda
k)+2k_2^2(1+2\lambda k)}{(1+\lambda k)(1+2\lambda k
)}\right]\,e^{-kh}\\
A_2^\parallel&=&\frac{2k_1 k_2}{hk^3}\left[\frac{ (1+\lambda k)(1+hk)+2\lambda
k}{(1+\lambda k)(1+2\lambda k )}\right]\,e^{-kh}, \\
A_3^\parallel &=&0, \\
B^\parallel& =&\frac{2 i k_1(hk-1)}{hk^2(1+2\lambda k)}\,e^{-kh}.
\end{subeqnarray}

These can be written as
\begin{equation}
A^\parallel_\alpha({\bf k},z,\lambda)=\frac{C^\parallel_\alpha({\bf
k},0)}{1+2\lambda k}+\frac{D^\parallel_\alpha({\bf
k},0)}{1+\lambda k}
\end{equation}
and
\be
B^\|({\bf k},z,\lambda)=\frac{B^\parallel({\bf
k},0)}{1+2\lambda k},
\ee
where 
\begin{equation}
C^\parallel_1({\bf k},z,0) = \frac{2k_1^2(hk-1)}{hk^3}e^{-kh},
\quad C^\parallel_2({\bf k},z,0) =  
\frac{ 2k_1 k_2(hk-1)}{hk^3}e^{-kh},
\end{equation}
\begin{equation}
D^\parallel_1({\bf k},z,0)=  -\frac{4k_2^2}
{hk^3}e^{-kh},\quad D^\parallel_2({\bf k},z,0)
= \frac{4k_1 k_2 }{hk^3}e^{-kh},
\end{equation}

As for the perpendicular case, we decompose the velocity field $\t {\bf w}^\parallel$ as
\begin{equation}
 \t\w^\parallel ({\bf r},\lambda) = \t\w_c^\parallel ({\bf r},\lambda)+
 \t\w_d^\parallel ({\bf r},\lambda)+
z\, \t\w_b^\parallel({\bf r},\lambda),
\end{equation}
where we now have
\begin{subeqnarray}
\left(1-2\lambda \frac{\p}{\p z}\right){\t{\bf
w}}_c^\parallel({\bf r},\lambda)&=&{\t{\bf w}}_c^\parallel({\bf
r},0),\slabel{int1_perp}
\\
\left(1-\lambda \frac{\p}{\p z}
\right){\t{\w}}_d^\parallel({\bf
r},\lambda)&=&{\t\w}_d^\parallel({\bf r},0),\slabel{int2_perp}
\\
\left(1-2\lambda \frac{\p}{\p z}
\right){{{{\t\w}}}}_b^\parallel({\bf
r},\lambda)&=&{{{\t\w}}}_b^\parallel({\bf r},0).
\slabel{int4_perp}
\end{subeqnarray}
so we get
\begin{equation}\label{parallel_tosolve}
\left(1-2\lambda \frac{\p}{\p z}\right)    \w^\parallel ({\bf r},\lambda) =  \w^\parallel ({\bf r},0)
- \lambda \frac{\p}{\p z}  \w_d^\parallel ({\bf r},\lambda)
-2\lambda  \w_b^\parallel({\bf r},\lambda),
\end{equation}
where $\w^\parallel ({\bf r},0)$ is such that
$\V^\parallel(\r)+\w^\parallel ({\bf r},0)$ is Blake's image system for a no-slip surface, that is
$\w^\parallel ({\bf r},0) = -2\G^S_x+2h\G^{SD}_{z;x}-2h^2\G^D_x $. Solving for each of the three terms of Eq.~\eqref{parallel_tosolve} using Eqs.~\eqref{int2_perp} and \eqref{int4_perp}  as well as integration by parts allows the flow field to be expressed as 
$\u^\parallel=\U^\parallel+\V^\parallel+\w^\parallel$, where
$\U^\parallel$ and $\V^\parallel$ are given by Eqs.~\eqref{Uparallel}
and \eqref{Vparallel}, and $\w^\parallel$ is given by
\begin{eqnarray}\label{imagepara2}
\w^\parallel(\r,\lambda) & = & \frac{1}{\lambda} \int_0^\infty 
[-\G^S_x+h\G^{SD}_{z;x}-h^2\G^{D}_x]({\bf  r} + (h+s){\bf e}_z)
\,e^{{-s}/{2\lambda}}\,\d s\nonumber\\
& + & {4\lambda} \int_0^\infty
 \g_d({\bf  r} + (h+s){\bf e}_z)\,\left[e^{{-s}/{2\lambda}} - 1\right]^2\,\d s \\
& + & 4  \lambda  \int_0^\infty \g_b ({\bf  r} + (h+s){\bf e}_z) \left[ 1-\left(1+\frac{s}{2\lambda}\right)\,e^{{-s}/{2\lambda}}
\right]
\,\d s, \nonumber
\end{eqnarray}
where we have defined two new singularities
\begin{equation}
\g_d ({\bf r})=\frac{1}{4} \frac{\p^2 }{\p z^2} [\w_d^\parallel({\bf r},0)],\quad \g_b({\bf r})=\frac{1}{2}\frac{\p }{\p z}[\w_b^\parallel ({\bf r},0)].
\end{equation}
Their Fourier transforms are given by
\begin{eqnarray}
\t \g_d({\bf k},z) & = & \frac{1}{8\pi\mu}\left(-\frac{k_2^2}{k},\frac{k_1k_2}{k} , 0 \right)e^{-kz}, \\
\t \g_b({\bf k},z) & = & \frac{(1-hk)}{8\pi\mu}\left(\frac{-k_1^2}{k},\frac{-k_1k_2}{k},ik_1 \right)e^{-kz},
\end{eqnarray}
and therefore, by inverse Fourier transforms, we find
\begin{eqnarray}
\label{gd}
\g_d({\bf r}) & = &\frac{\p}{\p y} \left[\frac{1}{8\pi\mu}  \left({\bf e}_z \times \frac{\bf r}{r^3}\right)\right]=-\G^{RD}_{z;y}({\bf r}),\\
\label{gb}
\g_b ({\bf r}) & = &\left( 1 + h \frac{\p}{\p z}\right)\left[ \G^{D}_x \right]=[\G^D_x-h \G^{Q}_{xz}]({\bf r})
\end{eqnarray}
where $\G^{RD}_{z;y}$ is the rotlet dipole in the $(y;z)$-direction, that is, the $y$-dipole of the $z$-rotlet $\G^R_z$, flow field due to a point torque and defined as
\begin{equation}
\G^{R}_{z}=\frac{1}{2}(\G^S_{y;x}-\G^S_{x;y}),
\end{equation}
and therefore
\begin{equation}
\G^{RD}_{z;y}=\frac{1}{2}(\G^{SD}_{y;xy}-\G^{SD}_{x;yy}).
\end{equation}

As a summary, the solution for $ \w^\parallel$ in the case of the parallel Stokeslet is given by
\begin{eqnarray}\label{imagepara_final}
\w^\parallel(\r,\lambda) & = & \frac{1}{\lambda} \int_0^\infty 
[-\G^S_x+h\G^{SD}_{z;x}-h^2\G^{D}_x]({\bf  r} + (h+s){\bf e}_z)
\,e^{{-s}/{2\lambda}}\,\d s\nonumber\\
& - & {4\lambda} \int_0^\infty \G^{RD}_{z;y}({\bf  r} + (h+s){\bf e}_z)\,\left[e^{{-s}/{2\lambda}} - 1\right]^2\,\d s \\
& + & 4  \lambda  \int_0^\infty [\G^D_x -h \G^{Q}_{xz} ]({\bf  r} + (h+s){\bf e}_z) \left[ 1-\left(1+\frac{s}{2\lambda}\right)\,e^{{-s}/{2\lambda}}
\right]
\,\d s, \nonumber
\end{eqnarray}
which is, again, a line integral of fundamental singularities distributed along a line in the $-{\bf e}_z$ direction, with a weighted magnitude. Unlike the perpendicular case, however, not all of the magnitudes of the singularities decay exponentially away from the image location.

\section{Some useful Fourier transforms}\label{Fourier}
Below is a list of two-dimensional Fourier transforms used to
derive the image systems in  Appendices \ref{solve_perpendicular} and \ref{solve_parallel}:
\begin{subeqnarray}\label{fourier2}
{\cal F}\left[\frac{1}{r} \right]=\frac{1}{k}\,e^{-kz} & , & \quad {\cal F}\left[\frac{z}{r^3} \right]=\,e^{-kz},\quad
{\cal F}\left[\frac{3xz}{r^5} \right]=ik_1\,e^{-kz},\\
{\cal F}\left[\frac{x}{r^3} \right]=\frac{ik_1}{k}\,e^{-kz}&,&\quad
{\cal F}\left[\frac{1}{r^3}\left(\frac{3z^2}{r^2}-1\right) \right]=k\,e^{-kz},\\
{\cal F}\left[\frac{3xy}{r^5}
\right]=-\frac{k_1k_2}{k}\,e^{-kz}&,& \quad
{\cal
F}\left[\frac{1}{r^3}\left(1-\frac{3x^2}{r^2}\right)\right]=\frac{k_1^2}{k}\,e^{-kz},
\\
{\cal F}\left[\frac{x^2}{r^3}
\right]= \frac{k_2^2-k_1^2zk}{k^3}\,e^{-kz}
&,&\quad {\cal F}\left[\frac{xy}{r^3}
\right]=-\frac{k_1k_2(1+zk)}{k^3}\,e^{-kz},
\end{subeqnarray}
where we have used the notation $r^2=x^2+y^2+z^2$.

%%%%%%%%%%% 						RECIPROCAL
\section{First influence of slip length on particle diffusivities: alternative method}
\label{recip}
We show in this section that the results given by Eq.~\eqref{change} for small slip length can also been obtained by using the reciprocal theorem. Let us consider the volume $V$ of fluid above the solid surface $S$ and two steady velocity fields, ${\bf u}$ and $\hat {\bf u}$, with corresponding stress tensors, $\ss$ and $\hat\ss$, and volume forcing  ${\bf f}$ and $\hat {\bf f}$ respectively. The reciprocal theorem states that
\begin{equation}
\int_S {\bf u}\cdot \hat\ss \cdot {\bf e}_z \,\d S-\int_S \hat{\bf u}\cdot \ss \cdot {\bf e}_z \,\d\,S= \int_V \hat{\bf f}\cdot{\bf u}\,\d V-\int_V {\bf f}\cdot\hat{\bf u}\,\d V.
\end{equation}
We take $\hat{\bf u}$ to be the flow field due to the point force $\hat{\bf F}$ at $(x,y,z)=(0,0,h)$ near a no-slip surface. Let us also consider the flow field, ${\bf v}$, due to the same point force near a surface with small slip length, in the sense $\lambda\ll h$. we perform a regular perturbation expansion and write 
\begin{equation}
{\bf v}=\hat{\bf u} + \kn \,{\bf u} + \oo{\kn^2}.
\end{equation}
The boundary condition for ${\bf v}$ on the surface is ${\bf v} = \lambda {\p {\bf v}}/{\p z}$, which becomes, at leading order in $\oo{\kn}$, ${\bf u} = h{\p \hat{\bf u}}/{\p z}$. Applying the reciprocal theorem for ${\bf u}$ and $\hat {\bf u}$ and using the relation between stress tensor and rate-of-strain tensor leads to
\begin{equation}\label{rec}
\mu h \int_S\left( \frac{\p \hat{\bf u}}{\p z}\cdot \frac{\p \hat{\bf u}}{\p z}\right) \,\d S= {\bf F}\cdot {\bf U},
\end{equation}
where $\mu$ is the shear viscosity of the fluid. In the case of a Stokeslet perpendicular to the surface, ${\bf F}=F_z{\bf e}_z$, Eq.~\eqref{rec} becomes
\begin{equation}
F_zU_z = \frac{9F^2_z}{2\pi\mu h}\int_0^\infty \frac{u^3}{(1+u^2)^5}\,\d u = \frac{9F_z^2}{48\pi \mu h}
\end{equation}
so 
\begin{equation}\label{rec_perp}
\frac{D^\perp}{D_0}= 1-\frac{9a}{8h}\left(1-\kn\right)\cdot
\end{equation}
In the case of a Stokeslet parallel to the surface ${\bf F}=F_x{\bf e}_x$, Eq.~\eqref{rec} becomes
\begin{equation}
F_xU_x = \frac{9 F^2_x}{4\pi\mu h}\int_0^\infty \frac{u^5}{(1+u^2)^5}\,\d u = \frac{9F_x^2}{96 \pi \mu h}
\end{equation}
so 
\begin{equation}\label{rec_parallel}
\frac{D^\parallel}{D_0}= 1-\frac{9a}{16h}\left(1-\kn\right)\cdot
\end{equation}
The results given by Eqs.~\eqref{rec_perp} and \eqref{rec_parallel} are the same as those obtained in Eq.~\eqref{change}.

%%%%%%%%%%% 						COUPLED
\section{Calculation of coupled mobilities}
\label{mobility_tensor}
For convenience, we define below the integrals
\begin{subeqnarray} \label{integrals}
{\bf I}_1 & = & \frac{1}{\lambda} \int_0^\infty  [-\G^S_x+h\G^{SD}_{z;x}-h^2\G^{D}_x]({\bf  r} + (h+s){\bf e}_z) \,e^{{-s}/{2\lambda}}\,\d s,\\
{\bf I}_2 & = & -{4\lambda} \int_0^\infty \G^{RD}_{z;y}({\bf  r} + (h+s){\bf e}_z)\,\left[e^{{-s}/{2\lambda}} - 1\right]^2\,\d s,\\
{\bf I}_3 & = & 4  \lambda  \int_0^\infty [\G^D_x -h \G^{Q}_{xz} ]({\bf  r} + (h+s){\bf e}_z) \left[ 1-\left(1+\frac{s}{2\lambda}\right)\,e^{{-s}/{2\lambda}}\right] \,\d s,
\end{subeqnarray}
so that the complete image system for a parallel Stokeslet, Eq.~\eqref{imagepara_final}, is written $\w^\parallel(\r,\lambda)={\bf I}_1+{\bf I}_2+{\bf I}_3$.

\subsection{Calculation of ${ b}_{x_2x_1}$}
Because of the decomposition in Eq.~\eqref{decomp}, the mobility is given by
\begin{equation}\label{decomp_bxx}
b_{x_2x_1}(\kn)=b_{x_2x_1}(\kn=\infty) + w_1^\parallel(d{\bf e}_x+h{\bf e}_z,\lambda).
\end{equation}
The integrals in Eq.~\eqref{integrals} and their asymptotic behaviors for small $\xi$ are given by
\begin{subeqnarray}
{\bf I}_1\cdot {\bf e}_x& = & \frac{1}{4\pi \mu d}\left[ \frac{1}{\kn}\int_0^\infty e^{-u/\kn} 
\left(\frac{\frac{3}{4}\xi(1+2u)}{\left[1+\xi(1+u)^2\right]^{5/2}}
-\frac{2+\xi(\frac{5}{4}+\frac{5}{2}u+u^2)}{\left[1+\xi(1+u)^2\right]^{3/2}}
 \right) \,\d u \right], \\
& = & \frac{1}{4\pi \mu d}\left[-2 + \xi\left(\frac{5}{2} + 5\kn + 4\kn^2\right) + \oo{\xi^2}\right], \\
{\bf I}_2\cdot {\bf e}_x & = & - \frac{1}{4\pi \mu d}\left[\kn \, \xi \int_0^\infty \frac{1}{[1+\xi(1+u)^2]^{3/2}}\left(e^{-u/\kn}-1\right)^2\,\d u \right], \\
& = & \frac{1}{4\pi \mu d} \left[ -\xi^{1/2}\kn + \xi \left(\kn+ \frac{3}{2}\kn^2\right) + \oo{\xi^2}\right],\\
{\bf I}_3\cdot {\bf e}_x  & = & \frac{1}{4\pi \mu d}\left[\kn \, \xi \int_0^\infty \left(\frac{2+\xi(\frac{1}{2}-\frac{1}{2}u-u^2)}{[1+\xi(1+u)^2]^{5/2}}-\frac{\frac{15}{2}\xi(1+u)}{[1+\xi(1+u)^2]^{7/2}}\right)\left(1-\left(1+\frac{u}{\kn}\right)e^{-u/\kn}\right)\,\d u \right], \\
& = & \frac{1}{4\pi \mu d} \left[\xi^{1/2}\kn -\xi(3\kn+4\kn^2) + \oo{\xi^2}\right],
\end{subeqnarray}
leading to the asymptotic behavior
\begin{equation}
w_1^\parallel(d{\bf e}_x+h{\bf e}_z,\lambda)=\frac{1}{4\pi \mu d} \left[-2 + \xi\left(\frac{5}{2}+3\kn+\frac{3}{2}\kn^2\right)+\oo{\xi^2}\right],
\end{equation}
and therefore, using Eqs.~\eqref{coupled_perfect} and \eqref{decomp_bxx}, to the mobility
\begin{equation}
b_{x_2x_1}(\kn)=\frac{3\xi}{8\pi \mu d}\left(1+2\kn + \kn^2\right)\,[1+\oo{\xi}].
\end{equation}
This asymptotic behavior is valid in the limit
$d \gg \max(h,\lambda)$.

\subsection{Calculation of ${ b}_{y_2y_1}$}

In this section, we suppose that the two particles are aligned along the $y$-axis, at a distance $d$ from each other, and apply a force in the $x$-direction on the first particle. The mobility obtained in this case is the component $b_{x_2x_1}$ for particles aligned along $y$, which is equal, by symmetry, to the $b_{y_2y_1}$ component for particles aligned along $x$.  Consequently, the mobility is given by
\begin{equation}\label{decomp_byy}
b_{y_2y_1}(\kn)=b_{y_2y_1}(\kn=\infty) + w_1^\parallel(d{\bf e}_y+h{\bf e}_z,\lambda).
\end{equation}
The integrals in Eq.~\eqref{integrals} and their asymptotic behaviors for small $\xi$ are now given by
\begin{subeqnarray}
{\bf I}_1\cdot {\bf e}_x & = &- \frac{1}{4\pi \mu d}\left[ \frac{1}{\kn}\int_0^\infty  \left( \frac{1+\xi(\frac{5}{4}+\frac{5}{2}u+u^2)}{\left[1+\xi(1+u)^2\right]^{3/2}} \right)\,e^{-u/\kn} \,\d u \right], \\
& = & \frac{1}{4\pi \mu d}\left[-1+\xi\left(\frac{1}{4}+\frac{1}{2}\kn+\kn^2\right)-\xi^2\left(\frac{3}{4}\kn^2+\frac{9}{2}\kn^3+9\kn^4\right) + \oo{\xi^3}\right], \\
{\bf I}_2\cdot {\bf e}_x & = & \frac{1}{4\pi \mu d}\left[\kn \, \xi \int_0^\infty \frac{2-\xi(1+u)^2}{[1+\xi(1+u)^2]^{5/2}}\left(e^{-u/\kn}-1\right)^2\,\d u \right] ,\\
& = & \frac{1}{4\pi \mu d} \left[
\xi^{1/2}\kn - \xi (2\kn+3\kn^2)+\xi^2 \left(2\kn + 9\kn^2 + 21\kn^3 + \frac{45}{2}\kn^4\right) + \oo{\xi^3}\right],\\
{\bf I}_3\cdot {\bf e}_x  & = & \frac{1}{4\pi \mu d}\left[\kn \, \xi \int_0^\infty \left(\frac{-1+\xi(\frac{1}{2}-\frac{1}{2}u-u^2)}{[1+\xi(1+u)^2]^{5/2}} \right)\left(1-\left(1+\frac{u}{\kn}\right)e^{-u/\kn}\right)\,\d u \right], \\
& = & \frac{1}{4\pi \mu d} \left[ - \xi^{1/2}\kn + \xi \left(\frac{3}{2}\kn + 2\kn^2\right) - \xi^2\left(\frac{5}{4}\kn + 6\kn^2 + \frac{27}{2}\kn^3 + 12 \kn^4 \right) + \oo{\xi^3}
\right],
\end{subeqnarray}
so that
\begin{equation}
w_1^\parallel(d{\bf e}_y+h{\bf e}_z,\lambda)=\frac{1}{4\pi \mu d} \left[
-1 +\frac{1}{4}\xi + \xi^2\left(
\frac{3}{4}\kn  + \frac{9}{4}\kn^2 + 3\kn^3 + \frac{3}{2}\kn^4
\right)+\oo{\xi^3}
\right],
\end{equation}
and therefore, using Eqs.~\eqref{coupled_perfect} and \eqref{decomp_byy}, the mobility is given by
\begin{equation}
b_{y_2y_1}(\kn)=\frac{3\xi^2}{64\pi \mu d}\left(1+4 \kn + 12\kn^2+16\kn^3+8\kn^4\right)\,[1+\oo{\xi}].
\end{equation}
This asymptotic behavior is valid in the limit
$d \gg \max(h,\lambda)$.

\subsection{Calculation of ${ b}_{x_2z_1}$ ($=-{ b}_{z_2x_1}$)}

In this case, we have
\begin{equation}\label{decomp_bxz}
b_{x_2z_1}(\kn)=b_{x_2z_1}(\kn=\infty) + w_1^\perp(d{\bf e}_x+h{\bf e}_z,\lambda),
\end{equation}
with
\begin{eqnarray}
w_1^\perp(\r,\lambda) & = &  \frac{1}{8 \pi \mu d}\left[\frac{\xi^{1/2}}{\kn} \int_0^\infty \frac{1+\xi(u^2+\frac{1}{2}u-\frac{1}{2})}{[1+\xi(1+u)^2]^{5/2}}\,e^{-u/\kn}\,\d u \right] ,\\
& = & \frac{1}{8\pi \mu d}\left[\xi^{1/2} - \xi^{3/2} \left(3+\frac{9}{2}\kn+3\kn^2 \right) + \oo{\xi^{5/2}}\right] ,
\end{eqnarray}
and therefore, using Eqs.~\eqref{coupled_perfect} and \eqref{decomp_bxz}, we obtain
\begin{equation}
b_{x_2z_1}(\kn)=-\frac{3\xi^{3/2}}{16\pi \mu d} \left(1+3\kn+2\kn^2 \right)\,[1+\oo{\xi}].
\end{equation}
This asymptotic behavior is valid in the limit
$d^2 \gg \max(h^2,\lambda^2)$.

\bibliographystyle{unsrt}
\bibliography{slip}

\end{document}